\documentclass[prl, twocolumn, floatfix, superscriptaddress, nofootinbib, nolongbibliography]{revtex4-2}
\usepackage{graphicx}
\usepackage{amsmath,amssymb,mathtools,bm,tikz,physics}
\usepackage[colorlinks=true, allcolors=purple]{hyperref}

\newcommand{\Ising}{\text{I}}
\newcommand{\bh}{\boldsymbol{\eta}}
\newcommand{\be}{\boldsymbol{e}}

\begin{document}

\title{Intrinsic Error Thresholds in Nearly Critical Toric Codes}

\affiliation{Department of Physics and Institute for Quantum Information and Matter, California Institute of Technology, Pasadena, California 91125, USA}
\affiliation{Department of Physics, Princeton University, Princeton, New Jersey 08544, USA}
\affiliation{Department of Physics, University of California, Berkeley, California 94720, USA}

\author{Zack Weinstein}
\email{zackmw@caltech.edu}
\affiliation{Department of Physics and Institute for Quantum Information and Matter, California Institute of Technology, Pasadena, California 91125, USA}
\affiliation{Department of Physics, University of California, Berkeley, California 94720, USA}

\author{Samuel J. Garratt}
\affiliation{Department of Physics, Princeton University, Princeton, New Jersey 08544, USA}
\affiliation{Department of Physics, University of California, Berkeley, California 94720, USA}

\date{\today}

\begin{abstract}
We study the protection of information in nearly critical topological quantum codes, constructed by perturbing topological stabilizer codes towards continuous quantum phase transitions. Our focus is on the transverse-field toric code subjected to local Pauli decoherence. Despite the strong quantum fluctuations of anyons when the transverse field is tuned infinitesimally close to the critical point, we show that a \emph{finite} strength of Pauli decoherence remains necessary to irreversibly destroy information encoded in the ground-state manifold. Using a replica statistical physics mapping for the coherent information, we show that decoherence can be understood as introducing a two-dimensional inter-replica defect within a three-dimensional replica statistical physics model. A field theoretical analysis shows that this defect is perturbatively irrelevant to the bulk critical point, and cannot renormalize the transverse field strength, leading to a finite error threshold. We argue that a qualitatively similar conclusion can be drawn for a broad class of nearly critical topological codes, under a variety of decoherence channels.
\end{abstract}

\maketitle

A defining feature of topological phases of matter is their capacity to robustly protect quantum information from decoherence \cite{kitaev_fault-tolerant_2003,dennis_topological_2002,bravyi_topological_2010}. This property has led to significant theoretical \cite{dassarma_topologically_2005,bravyi_universal_2006,freedman_towards_2006,alicea_new_2012,tantivasadakarn_shortest_2022,tantivasadakarn_hierarchy_2022} and experimental \cite{satzinger_realizing_2021,semeghini_probing_2021,leonard_realization_2023,iqbal_topological_2024,iqbal_non-abelian_2024} interest in realizing topologically ordered states as platforms for scalable quantum error correction \cite{kitaev_anyons_2006,nayak_nonabelian_2008}.

While most theoretical studies focus on the protection of information in exactly solvable models \cite{dennis_topological_2002,wang_confinement-higgs_2003,katzgraber_error_2009,stace_thresholds_2009,bombin_strong_2012,fan_diagnostics_2024,bao_mixedstate_2023,leeQuantumCriticalityDecoherence2023,sangMixedStateQuantumPhases2024,chen_unconventional_2024,chenSeparabilityTransitionsTopological2024,suTapestryDualitiesDecohered2024,lyons_understanding_2024,wang_intrinsic_2025,wang_fractional_2025,temkin_charge-informed_2025}, it is natural to ask how the protection of quantum information is degraded as these idealized models are deformed away from their solvable limits. For sufficiently weak Hamiltonian perturbations, topological order in these models is stable \cite{hastings_quasiadiabatic_2005,bravyi_topological_2010,bravyi_short_2011,michalakis_stability_2013,nachtergaele_stability_2024} and defines an entire topological phase, which is expected to robustly protect quantum information deep within the phase. It is therefore especially interesting to ask how, or if, this protection breaks down as these models are pushed towards a quantum phase transition to a trivial phase with no information-encoding capacity.

\begin{figure}[b]
    \includegraphics[width = \columnwidth]{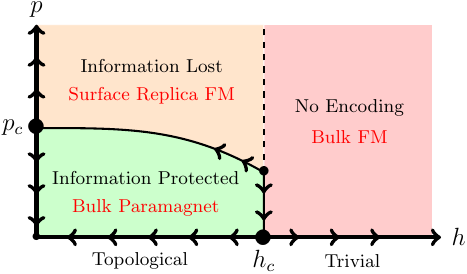}
    \caption{Phases of information protection in the transverse-field toric code (TFTC) at field strength $h$, subjected to bit-flip decoherence of strength $p$. For all $h < h_{c}$, the TFTC exhibits a topological phase with four (nearly) degenerate ground states, which can be used to define a quantum code. The critical strength of bit-flip decoherence necessary to destroy the encoded information is finite throughout the phase, and remains finite even as $h \to h_{c}$. Red text denotes the phases of the 3$d$ replica statistical physics model which describes the R\'enyi coherent information; for $h < h_{c}$, information is destroyed when a 2$d$ defect surface exhibits an ordering transition.}
    \label{fig:frontpage}
\end{figure}

In this Letter, we study intrinsic error thresholds in topological quantum codes as they are pushed towards quantum critical points. Our focus is primarily on the transverse-field toric code (TFTC) \cite{wegnerDualityGeneralizedIsing1971,kogutIntroductionLatticeGauge1979,sachdevQuantumPhaseTransitions2011,fradkin_field_2013,trebst_breakdown_2007}, where the transverse field provides quantum fluctuations to the $m$ anyons. When these anyons are driven towards condensation at a critical field strength, a natural expectation is that the resulting codes become highly unstable to decoherence which further proliferates the anyons. We show that this naive intuition is incorrect: in fact, we demonstrate that the intrinsic threshold to bit-flip decoherence remains \emph{finite} as the critical point is approached (see Fig.~\ref{fig:frontpage}).

\begin{figure}[t]
    \includegraphics[width = 0.75\columnwidth]{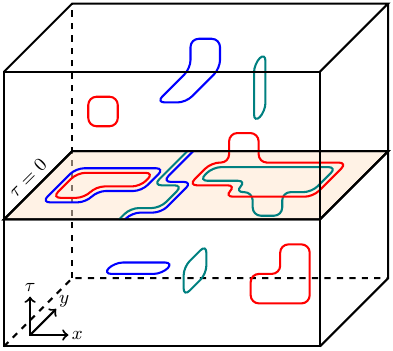}
    \caption{Depiction of the replica statistical physics model~\eqref{eq:purity_statmech}, which is used to compute the R\'enyi coherent information $I_{c}^{(n)}$ [Eq.~\eqref{eq:renyi_coherent_info}]. Each $n$th replica model consists of $n$ 3$d$ Ising models, represented using a high-temperature expansion as $n$ species of closed-loop anyon worldlines. The decoherence favors correlated worldline trajectories within the $\tau = 0$ surface. When pairs of worldlines proliferate along this surface, $I_{c}^{(n)}$ transitions from $2 - \mathcal{O}(e^{-L/\xi})$ to $0 + \mathcal{O}(e^{-L/\xi})$.}
    \label{fig:replicas}
\end{figure}

Our theoretical evidence for this claim builds on previous works studying the effects of measurements \cite{garrattMeasurementsConspireNonlocally2023a,weinsteinNonlocalityEntanglementMeasured2023a,yang_entanglement_2023,murciano_measurement_2023} and decoherence \cite{leeQuantumCriticalityDecoherence2023,zouChannelingQuantumCriticality2023,bao_mixedstate_2023} on both critical and topological quantum ground states. We specifically analyze the coherent information, a standard theoretical diagnostic for the preservation of quantum information under decoherence \cite{schumacherQuantumDataProcessing1996,schumacher_approximate_2002,fan_diagnostics_2024}, as the $n \to 1$ replica limit of the $n$th R\'enyi coherent information. Utilizing a duality transformation \cite{wegnerDualityGeneralizedIsing1971} and the quantum-classical mapping \cite{fradkin_order_1978,kogutIntroductionLatticeGauge1979,sachdevQuantumPhaseTransitions2011,fradkin_field_2013}, we show that decoherence can be understood as coupling $n$ replicas of the 3$d$ Ising model by their energy densities along a single 2$d$ `defect surface' (Fig.~\ref{fig:replicas}). In particular, the R\'enyi coherent information is mapped to the difference of free energies in the replica model for two different spatial boundary conditions. This quantity transitions from its maximal value to zero when the defect surface orders at a critical decoherence threshold.

As the transverse field $h$ approaches its critical strength $h_c$, we can utilize a coarse-grained replica field theory to determine the general structure of the phase diagram (Fig.~\ref{fig:frontpage}). The renormalization group (RG) flow in the vicinity of the `bulk' critical point is constrained by two key features: first, the decoherence-induced inter-replica coupling is perturbatively irrelevant at the bulk critical point; second, this surface coupling cannot renormalize the bulk couplings of the Ising models \cite{diehl_fieldtheoretical_1981,cardy_conformal_1984,cardyScalingRenormalizationStatistical1996,diehl_theory_1997,metlitski_boundary_2022,krishnan_plane_2023}. These features imply that a finite strength of decoherence is required to induce a surface-ordering transition and degrade the coherent information, even near the critical point. The resulting phase diagram is depicted in Figure~\ref{fig:frontpage}.

While the decohered TFTC admits an especially simple microscopic analysis, our field-theoretic conclusions apply to a broad class of Hamiltonian perturbations to the toric code. This includes the Fradkin-Shenker model, i.e., the toric code under both transverse and longitudinal fields \cite{fradkinPhaseDiagramsLattice1979,tupitsyn_topological_2010,wu_phase_2012,verresenHiggsCondensatesAre2022}. In the End Matter, we provide an effective field-theoretic analysis which generalizes our TFTC results to generic transitions out of the toric code's topological phase driven by a single condensing boson.

In a companion work \cite{lip}, we study explicit decoding protocols for the TFTC under bit-flip decoherence. There we numerically demonstrate a decoder whose error threshold approaches a finite value as the TFTC is tuned towards criticality, providing a direct confirmation of the present work's results. We additionally prove that \emph{phase-flip} decoherence, which incoherently proliferates the gapped $e$ anyons, exhibits a \emph{constant} error threshold throughout the topological phase of the TFTC.

\emph{Model.}---  We study intrinsic error thresholds in the TFTC on an $L \times L$ periodic square lattice. The Hamiltonian is $H \equiv - \sum_{v} A_{v} - \sum_{p} B_{p} - h \sum_{\ell} X_{\ell}$, where $X_{\ell}$ and $Z_{\ell}$ are Pauli operators for qubits on the links of the lattice, and $A_{v} \equiv \prod_{\ell \ni v} X_{\ell}$ and $B_{p} \equiv \prod_{\ell \in p} Z_{\ell}$ are the standard vertex and plaquette stabilizers. For $h < h_{c} \approx 0.328$ \cite{wu_phase_2012}, $H$ exhibits a four-dimensional, nearly degenerate manifold $\mathcal{C}_{h}$ of topologically ordered ground states \cite{sachdevQuantumPhaseTransitions2011}, which can encode two qubits of quantum information. The transverse field $h$ induces quantum fluctuations of $m$ anyon excitations $B_{p} = -1$, which condense and destroy the topological order for $h \geq h_{c}$.

We are interested in the protection of quantum information in the family of quantum codes $\mathcal{C}_{h}$, especially as $h$ approaches $h_{c}$. Here we focus on i.i.d. bit-flip decoherence of strength $p$, described by the channel $\mathcal{E}_{\ell}(\rho) \equiv (1 - p) \rho + p X_{\ell} \rho X_{\ell}$ acting on each link $\ell$ of the lattice. This channel can be understood as incoherently creating a pair of $m$ anyons on the two plaquettes adjacent to $\ell$, with probability $p$. The total channel is $\mathcal{E} \equiv \prod_{\ell} \mathcal{E}_{\ell}$. 

To determine whether there \emph{in principle} exists a recovery channel which can invert $\mathcal{E}$ on $\mathcal{C}_{h}$ with high fidelity, we analyze the coherent quantum information~\cite{schumacherQuantumDataProcessing1996,schumacher_approximate_2002}. Let us denote the principal system as $Q$, and introduce a two-qubit reference system $R$. We begin with a maximally entangled state $\ket{\Psi_{QR}}$ between the reference $R$ and the ground-state manifold $\mathcal{C}_{h}$ in $Q$ [see Eq.~\eqref{eq:psi_QR} in the End Matter]. After subjecting $Q$ to bit-flip decoherence, we obtain the mixed states $\rho_{QR} \equiv \mathcal{E}(\dyad{\Psi_{QR}})$ and $\rho_{Q} \equiv \tr_{R} \rho_{QR}$ \cite{fn1}. The coherent information is then defined as $I_{c}(R \rangle Q) \equiv S(\rho_{Q}) - S(\rho_{QR})$, where $S(\rho) \equiv -\tr [\rho \log_{2} \rho]$ is the von Neumann entropy of $\rho$. A near-maximal coherent information $I_{c}(Q \rangle R) \geq 2 - \varepsilon$ implies the existence of a recovery channel on $Q$ which recovers $\ket{\Psi_{QR}}$ with fidelity at least $1 - 2\sqrt{\varepsilon}$ \cite{schumacher_approximate_2002}.

In the stabilizer limit $h = 0$, the coherent information can be directly related to the free energy cost of imposing twisted boundary conditions in the 2$d$ Nishimori random-bond Ising model at disorder strength $p$ \cite{dennis_topological_2002,fan_diagnostics_2024,lee_exact_2025}. Although $I_{c}$ cannot be calculated as directly for $h > 0$, we can gain significant theoretical insight by studying the R\'enyi coherent information,
\begin{equation}
    \label{eq:renyi_coherent_info}
    I_{c}^{(n)}(R \rangle Q) \equiv \frac{1}{1 - n} \log_{2} \qty{ \frac{\tr \rho_{Q}^{n}}{\tr \rho_{QR}^{n}} } .
\end{equation}
In principle, $I_{c}$ can be computed from $I_{c}^{(n)}$ via the replica limit $n \to 1$.

\emph{Statistical Physics Mapping.}--- To determine the behavior of $I_{c}^{(n)}$ as a function of the transverse field $h$ and decoherence strength $p$, we will express the $n$th purities $\tr \rho_{Q}^{n}$ and $\tr \rho_{QR}^{n}$ as partition functions for an $n$-fold replicated statistical physics model, whose replicas are coupled by the decoherence. Towards this end, let $\ket{\psi_{h}^{\bh}}$ with $\bh \equiv (\eta_{x}, \eta_{y})$ and $\eta_{x}, \eta_{y} = \pm 1$ denote the four lowest-energy eigenstates of $H$. These are eigenstates of the two independent logical-$X$ operators $\overline{X}_{x}, \overline{X}_{y}$, satisfying $\overline{X}_{x} \ket{\psi^{\bh}_{h}} = \eta_{x} \ket{\psi^{\bh}_{h}}$ and $\overline{X}_{y} \ket{\psi^{\bh}_{h}} = \eta_{y} \ket{\psi^{\bh}_{h}}$. As shown in the End Matter, the purities $\tr \rho_{Q}^{n}$ and $\tr \rho_{QR}^{n}$ can be straightforwardly expressed in this basis as
\begin{subequations}
    \label{eq:purities}
    \begin{align}
        \tr \rho_{Q}^{n} &= \frac{1}{4^{n}} \sum_{\bh} \bra{\psi^{\bh \ldots \bh}_{h}} \hat{\mathcal{E}}^{(n)} \ket{\psi^{\bh \ldots \bh}_{h}} , \label{eq:rho_Q} \\
        \tr \rho_{QR}^{n} &= \frac{1}{4^{n}} \sum_{\bh_{1} \ldots \bh_{n}} \bra{\psi^{\bh_{1} \ldots \bh_{n}}_{h}} \hat{\mathcal{E}}^{(n)} \ket{\psi_{h}^{\bh_{1} \ldots \bh_{n}}} , \label{eq:rho_QR}
    \end{align}
\end{subequations}
where $\ket{\psi^{\bh_{1} \ldots \bh_{n}}_{h}} \equiv \ket{\psi^{\bh_{1}}_{h}} \otimes \ldots \otimes \ket{\psi^{\bh_{n}}_{h}}$, and the $n$th \emph{defect operator} $\hat{\mathcal{E}}^{(n)}$ is given by
\begin{equation}
    \label{eq:defect}
    \hat{\mathcal{E}}^{(n)} \equiv \exp \qty{ \gamma \sum_{\ell} \sum_{\alpha = 1}^{n} \qty[ X_{\ell}^{(\alpha)} X_{\ell}^{(\alpha+1)} - 1 ] } .
\end{equation}
Here $X_{\ell}^{(\alpha)}$ is the Pauli-$X$ operator acting on the $\alpha$th tensor factor, $\gamma \equiv \tanh^{-1}[p / (1-p)]$, and $X_{\ell}^{(n + 1)} \equiv X_{\ell}^{(1)}$. 

Since the operator $X_{\ell}^{(\alpha)}$ creates and hops $m$ anyons in the $\alpha$th replica (denoted $m^{(\alpha)}$), we can interpret $\hat{\mathcal{E}}^{(n)}$ as a wavefunction deformation which attempts to condense the inter-replica anyon pair $m^{(\alpha)} m^{(\alpha + 1)}$ \cite{rokhsar_superconductivity_1988,castelnovo_quantum_2005,castelnovo_quantum_2008}. As we shall see, this condensation of anyon pairs is responsible for the degradation of the R\'enyi coherent information. Maximal decoherence $p \to \frac{1}{2}$ corresponds to $\gamma \to \infty$, whereupon $\hat{\mathcal{E}}^{(n)}$ becomes a projector which strictly locks $X_{\ell}^{(\alpha)} = X_{\ell}^{(\alpha')}$ for each $\alpha, \alpha' = 1, \ldots , n$. As a sanity check, the limit $\gamma = 0$ returns $\tr \rho_{Q}^{n} = 4^{-(n-1)}$ and $\tr \rho_{QR}^{n} = 1$ as expected, while the limit $\gamma \to \infty$ demands that $\bh_{1} = \ldots = \bh_{n}$ in Eq.~\eqref{eq:rho_QR}, resulting in $I_{c}^{(n)} = 0$.

Away from these extreme limits, we can determine the behavior of $I_{c}^{(n)}$ by expressing the quantities $\mathcal{Z}^{(\bh_{1} \ldots \bh_{n})} \equiv \bra{\psi^{\bh_{1} \ldots \bh_{n}}_{h}} \hat{\mathcal{E}}^{(n)} \ket{\psi^{\bh_{1} \ldots \bh_{n}}_{h}}$ as partition functions of classical statistical physics models. We leave technical details for the Supplemental Material \cite{som}, and simply outline the result here. First, it is convenient to represent $\ket{\psi^{\bh}_{h}} \propto e^{-\beta H / 2} \ket{\psi^{\bh}_{0}}$ using imaginary-time evolution from the corresponding stabilizer toric code state $\ket{\psi^{\bh}_{0}}$. By trotterizing this imaginary-time evolution, and utilizing Wegner duality \cite{wegnerDualityGeneralizedIsing1971}, the matrix element $\bra{\psi^{\bh}_{0}} e^{-\beta H} \ket{\psi^{\bh}_{0}}$ can be represented as the partition function of an extremely anisotropic 3$d$ Ising model:
\begin{subequations}
    \label{eq:qc_mapping}
    \begin{align}
        \bra{\psi^{\bh}_{0}} e^{-\beta H} \ket{\psi^{\bh}_{0}} &\propto \sum_{\qty{\sigma}} e^{-S^{\bh}_{\Ising}[\sigma]} , \label{eq:matrix_elmt} \\
        S^{\bh}_{\Ising}[\sigma] &\equiv - \sum_{\expval{ij}} J^{\bh}_{ij} \sigma_{i} \sigma_{j} .
    \end{align}
\end{subequations}
Here $\sigma_{i} = \pm 1$ are Ising spins on the sites $i$ of an $L \times L \times M$ spacetime lattice, with $J_{ij}^{\bh} = \varepsilon h U^{\bh}_{ij}$ on spacelike bonds and $J_{ij}^{\bh} = \frac{1}{2} \log \frac{1}{\varepsilon}$ on timelike bonds, and $\varepsilon \equiv \frac{\beta}{M}$ is the size of a Trotter step. The partition function in Eq.~\eqref{eq:qc_mapping} exhibits open boundary conditions in the imaginary time direction, and periodic (antiperiodic) boundary conditions in the $x$ direction for $\eta_{x} = +1$ ($\eta_{x} = -1$), and similarly in the $y$ direction. This is implemented by a non-dynamical $\mathbb{Z}_{2}$ gauge field $U_{ij}^{\bh} = \pm 1$. We can regard the closed loops arising from the high-temperature expansion \cite{kardarStatisticalPhysicsFields2007} of Eq.~\eqref{eq:matrix_elmt} as the worldlines of $m$ anyons in Euclidean spacetime (Fig.~\ref{fig:replicas}), which exhibit finite line tension for $h < h_{c}$ and proliferate as $h \to h_{c}$. The topological phase of the TFTC is therefore described by the paramagnetic phase of the 3$d$ Ising model \cite{kogutIntroductionLatticeGauge1979,sachdevQuantumPhaseTransitions2011,fradkin_field_2013}.

In a similar manner, $\mathcal{Z}^{(\bh_{1} \ldots \bh_{n})}$ is represented in terms of $n$ such Ising models, cyclically coupled along a single imaginary time slice $\tau = 0$ by decoherence:
\begin{subequations}
    \label{eq:purity_statmech}
    \begin{align}
        &\mathcal{Z}^{(\bh_{1} \ldots \bh_{n})} \propto \sum_{\qty{\sigma}} \exp \qty{ - \sum_{\alpha = 1}^{n} S_{\Ising}^{\bh_{\alpha}}[\sigma^{(\alpha)}] - \delta S[\sigma] } , \label{eq:partition_fn} \\
        &\delta S[\sigma] \equiv - \gamma \sum_{\expval{i j} | \tau = 0} \sum_{\alpha = 1}^{n} (\sigma_{i} U^{\bh}_{ij} \sigma_{j})^{(\alpha)} (\sigma_{i} U^{\bh}_{ij} \sigma_{j})^{(\alpha + 1)} , \label{eq:defect_statmech}
    \end{align}
\end{subequations}
where $\sigma_{i}^{(\alpha)}$ is the Ising spin at site $i$ of the $\alpha$th replica, the first sum in Eq.~\eqref{eq:defect_statmech} is performed over bonds within the $\tau = 0$ imaginary time slice, and $(\sigma_{i} U_{ij}^{\bh} \sigma_{j})^{(\alpha)} \equiv \sigma_{i}^{(\alpha)} U_{ij}^{\bh_{\alpha}} \sigma_{j}^{(\alpha)}$ is the local energy density in the $\alpha$th replica. This inter-replica coupling reduces the free energy of $m$ anyon worldlines from adjacent replicas which traverse the $\tau = 0$ surface in a correlated fashion (Fig.~\ref{fig:replicas}). For $\gamma$ beyond a critical strength $\gamma^{(n)}_{c}(h)$, pairs of anyon worldlines proliferate within this surface. In terms of the Ising spins, this corresponds to a surface ordering transition in which the order parameter $\sigma_{i}^{(\alpha)} \sigma_{i}^{(\alpha')}$ develops long-range correlations for sites $i$ within the $\tau = 0$ surface. We refer to this phase as a `surface replica ferromagnet'.

The purities \eqref{eq:purities} are both proportional to the partition function of the same replica statistical physics model, with slightly different boundary conditions. In $\tr \rho_{Q}^{n}$, the boundary conditions of all $n$ replicas are locked together, whereas in $\tr \rho_{QR}^{n}$ the boundary conditions of each replica are allowed to fluctuate independently. $I_{c}^{(n)}$ is therefore proportional to the free energy cost of locking the replicas' boundary conditions together. The phase diagram of this replica model, and the behavior of $I_{c}^{(n)}$ within each phase, then follows from simple Landau theory:
\begin{enumerate}
    \item For $h < h_{c}$ and $\gamma < \gamma_{c}^{(n)}(h)$, the model is paramagnetic. The partition functions \eqref{eq:partition_fn} are largely independent of their boundary conditions $\qty{\bh_{\alpha}}$, and $\tr \rho_{QR}^{n} \simeq 4^{n-1}\tr \rho_{Q}^{n}$ up to exponentially small corrections in $L$. Consequently, $I_{c}^{(n)} = 2 - \mathcal{O}(e^{-L/\xi})$.

    \item For $h < h_{c}$ and $\gamma > \gamma_{c}^{(n)}(h)$, we obtain a surface replica ferromagnet \cite{fn5}. Boundary conditions in which $\bh_{\alpha} \neq \bh_{\alpha'}$ for some $\alpha \neq \alpha'$ exhibit a non-contractible domain wall in the order parameter $\sigma^{(\alpha)}_i \sigma^{(\alpha')}_i$, which costs an $\mathcal{O}(L)$ free energy. As a result, $\tr \rho_{QR}^{n} \simeq \tr \rho_{Q}^{n}$, and $I_{c}^{(n)} = 0 + \mathcal{O}(e^{-L/\xi})$.
    
    \item For $h > h_{c}$, we obtain ferromagnetic order throughout the $\tau \neq 0$ bulk and the $\tau = 0$ surface. For any infinitesimal $\gamma > 0$, Eq.~\eqref{eq:defect_statmech} again exhibits an $\mathcal{O}(L)$ free energy cost for boundary conditions in which $\bh_{\alpha} \neq \bh_{\alpha'}$ for some $\alpha \neq \alpha'$. Similar to the previous case, $I_{c}^{(n)} = 0 + \mathcal{O}(e^{-L/\xi})$.    
\end{enumerate}
In summary, we have a simple qualitative picture for the destruction of R\'enyi coherent information in each $n$th replica model: the effect of decoherence is to introduce an inter-replica surface defect at imaginary time $\tau = 0$, which couples each Ising model by their energy densities. $I_{c}^{(n)}$ is sharply reduced from its maximal value to zero when this surface exhibits an ordering transition. 

In the stabilizer limit $h = 0$, the $\tau \neq 0$ bulk spins can be trivially integrated out, and we obtain an effectively 2$d$ model described by Eq.~\eqref{eq:defect_statmech} alone. This precisely reproduces the replica models for the decohered toric code studied in Refs.~\cite{fan_diagnostics_2024,bao_mixedstate_2023,leeQuantumCriticalityDecoherence2023} (up to a minor change of variables). More generally, deep within the topological phase $h \ll h_{c}$, the bulk spins are paramagnetic and can (in principle) be integrated out to create renormalized short-range couplings within the $\tau = 0$ surface. If this qualitative picture can be extended to the replica limit $n \to 1$, it would suggest that the decoherence-induced destruction of coherent information is described by a continuous phase transition in a 2$d$ universality class, and that this transition arises at a finite value of the decoherence strength $p_{c}(h) > 0$ for $h \ll h_c$.

\emph{Replica Field Theory.}--- As $h \to h_{c}$, the $m$ anyons are on the verge of condensing within the ground states of $H$. One might expect that any further proliferation of anyons by decoherence will immediately trigger their condensation, destroying the encoded information. Here we will show that this is not the case: using an effective replica field theory in the vicinity of $h_{c}$, we will show that $p_{c}(h)$ asymptotes to a \emph{finite} value as $h \to h_{c}$.

Since the bulk correlation length diverges as $h \to h_{c}$, we can trade our lattice model \eqref{eq:partition_fn} for a continuum description. Each replica is described by a 3$d$ Ising* conformal field theory (CFT) \cite{fn2} with a scalar order parameter $\phi^{(\alpha)}(\tau, x)$, with $\alpha = 1, \ldots , n$, which roughly describes the condensing anyon $m^{(\alpha)}$. As we have seen, the decoherence functions as an inter-replica interaction which cyclically couples the replicas by their energy densities $\varepsilon^{(\alpha)}(\tau, x) \sim [\phi^{(\alpha)}(\tau, x)]^{2}$. Therefore, we expect that the $n$th replica model can be described at long wavelengths by an action of the form \cite{fn6}
\begin{subequations}
\label{eq:S}
    \begin{align}
        \mathcal{S}[\qty{\phi}] &= \sum_{\alpha = 1}^{n} \mathcal{S}_{0}[\phi^{(\alpha)}] + \delta \mathcal{S}[\qty{\phi}] + \ldots, \label{eq:S_tot} \\
        \mathcal{S}_{0}[\phi^{(\alpha)}] &\equiv \mathcal{S}_{\Ising}[\phi^{(\alpha)}] + r \int \dd{\tau} \dd[2]{x} \varepsilon^{(\alpha)}(\tau, x) , \label{eq:S_bulk} \\
        \delta \mathcal{S}[\qty{\phi}] &\equiv - \tilde{\gamma} \int \dd[2]{x} \sum_{\alpha = 1}^{n} \varepsilon^{(\alpha)}(0, x) \varepsilon^{(\alpha + 1)}(0, x) . \label{eq:S_defect}
    \end{align}
\end{subequations}
Here $\mathcal{S}_{\Ising}[\phi^{(\alpha)}]$ formally represents the Ising* CFT in replica $\alpha$, $r \propto h_{c} - h$ pushes the system towards its paramagnetic phase, and $\tilde{\gamma} \propto \gamma \propto p$ at small decoherence strengths. The ellipsis in \eqref{eq:S_tot} denotes less relevant terms. Near the critical point, the R\'enyi coherent information is given by the free energy cost of locking the replicas' boundary conditions in the above field theory.

To determine the effect of weak decoherence near the critical point, we compute the perturbative RG flow near the fixed point $(r, \tilde{\gamma}) = (0, 0)$. To leading nontrivial order, the most general RG equations are given by
\begin{subequations}
    \label{eq:rg}
    \begin{align}
        \dv{r}{\ell} &= (3 - \Delta_{\varepsilon}) r + \mathcal{O}(r^{2}) , \\
        \dv{\tilde{\gamma}}{\ell} &=  \qty[ (2 - 2 \Delta_{\varepsilon}) + c_{1} r + c_{2} \tilde{\gamma} + \mathcal{O}(r^{2}, \tilde{\gamma}^{2}) ] \tilde{\gamma} ,
    \end{align}
\end{subequations}
where $\Delta_{\varepsilon} \approx 1.4$ is the scaling dimension of the energy operator in the 3$d$ Ising CFT \cite{hasenbusch_finite_2010,kos_bootstrapping_2014}, while $c_{1}$ and $c_{2}$ are (possibly $n$-dependent) undetermined constants. Note first that the parameter $r$ is a bulk variable while $\tilde{\gamma}$ is a surface variable, and consequently $r$ cannot be renormalized by $\tilde{\gamma}$ \cite{cardyScalingRenormalizationStatistical1996}. Second, note that $\tilde{\gamma}$ is perturbatively irrelevant. Finally, note that the leading contributions to the RG equations are $n$-independent, allowing for a straightforward analytical continuation of the RG equations to the $n \to 1$ replica limit of interest.

Taken together, these facts imply that the perturbative RG flow in the $(r, \tilde{\gamma})$ plane must point vertically downwards towards $\tilde{\gamma} = 0$ along the $r = 0$ axis. This immediately leads to a finite value of the critical inter-replica coupling $\tilde{\gamma}_{c}(r)$ as $r \to 0$ \cite{fn4}. Recalling that $\tilde{\gamma} \propto p$, we have therefore shown that the intrinsic error threshold $p_c(h)$ remains finite as $h \to h_c$.

\emph{Discussion.}--- In this Letter, we have shown that topological quantum codes can robustly protect quantum information even as they are driven towards a quantum phase transition into a trivial phase. Even for anyons on the verge of condensation, such near-critical topological codes retain a finite intrinsic threshold to decoherence which further proliferates these anyons. While the TFTC under bit-flip decoherence served as a particularly simple case study, the physical picture we have developed applies quite broadly to many continuous phase transitions driven by condensing self-bosons, examples of which are abundant for both abelian and non-abelian topological orders \cite{bais_condensate_2009,burnell_anyon_2018}. In the End Matter, we provide an effective field theory argument which generalizes the results of this Letter to a broad class of nearly critical $\mathbb{Z}_2$ and $\mathbb{Z}_q$ topological codes under local Pauli decoherence.

From a dual perspective, the present work also has implications for strong-to-weak spontaneous symmetry breaking (SWSSB) \cite{leeQuantumCriticalityDecoherence2023,lessa2024strongtoweakspontaneoussymmetrybreaking,sala2024spontaneousstrongsymmetrybreaking,weinstein_efficient_2025} in near-critical symmetric phases under decoherence. In the Supplemental Material~\cite{som}, we discuss how our results on the TFTC can be reinterpreted in terms of the near-critical (2+1)$d$ transverse-field Ising model under $\mathbb{Z}_{2}$-symmetric decoherence. Analogous to the preceding discussion, we show that a finite strength of decoherence is required to induce SWSSB even as the transition to the ferromagnetic phase is approached.

A notable drawback of the present work is its reliance on the replica trick, and in particular the analytical continuation of the RG equations \eqref{eq:rg} to the replica limit $n \to 1$. Indeed, we cannot rigorously rule out the possibility that the replica limit fails to correctly describe the von Neumann coherent information. Additionally, while information is \emph{in principle} preserved below the intrinsic decoherence threshold, it is a priori unclear whether this information can be recovered by any efficient or practical decoding algorithm. In a companion work \cite{lip}, we resolve both of these issues by constructing and investigating an efficient optimal decoder for the TFTC under independent bit-flip and phase-flip decoherence. There we show that the decoder exhibits a finite threshold near criticality which is only slightly reduced from its value in the stabilizer toric code.

\makeatletter
\let\old@addcontentsline\addcontentsline
\renewcommand{\addcontentsline}[3]{}
\makeatother

\begin{acknowledgments}
    \emph{Acknowledgments.}--- We thank Ehud Altman for helpful feedback on the manuscript, and Ruihua Fan, Steve Simon, and Simon Trebst for fruitful discussions. Z.W. acknowledges funding provided by the Institute for Quantum Information and Matter, an NSF Physics Frontiers Center (NSF Grant PHY-2317110). S.J.G.~was supported by the Gordon \& Betty Moore Foundation, and in part by a Brown Investigator Award, a program of the Brown Institute for Basic Sciences at the California Institute of Technology. 
\end{acknowledgments}
\addcontentsline{toc}{section}{}

\bibliographystyle{apsrev4-2-author-truncate}
\bibliography{refs}

\begin{thebibliography}{95}
\bibitem [{fn5()}]{fn5}%
  \BibitemOpen
  \href@noop {} {}\bibinfo {note} {In principle, there can exist a fourth `surface ferromagnet' phase in which $\sigma^{(\alpha)}_i$ orders only within the $\tau = 0$ surface. Such a phase exhibits exponentially small R\'enyi coherent information, but also exhibits long-range correlations in $\tr[T_P \rho_Q^n] / \tr \rho_Q^n $, where $T_P \equiv \prod_{\ell \in P} X_{\ell}$ creates $m$ anyons at the endpoints of the dual-lattice string $P$. In the $n \to 1$ replica limit, $\tr[T_P \rho_Q] = \frac{1}{4} \sum_{\bh} \bra{\psi^{\bh}_h} T_P \ket{\psi^{\bh}_h}$ decays exponentially for all $h < h_c$. Consequently, this phase cannot possibly survive the replica limit.}\BibitemShut {Stop}%
\bibitem [{fn4()}]{fn4}%
  \BibitemOpen
\bibfield  {number} {  }\href@noop {} {}\bibinfo {note} {For example, if the coefficient $c_{2}$ happens to be positive and large compared to $2\Delta_{\varepsilon} - 2$, then we can truncate the RG flow at second order in $\tilde{\gamma}$ to obtain a critical inter-replica coupling $\tilde{\gamma}_{c}(r = 0) \simeq (2 \Delta_{\varepsilon} - 2) / c_{2}$.}\BibitemShut {Stop}%
\end{thebibliography}
\addcontentsline{toc}{section}{}

\newpage
\onecolumngrid
\section{End Matter}
\twocolumngrid

\emph{Appendix A: Deriving Eq.~\eqref{eq:purities}.}--- Here we show how the purities $\tr \rho_Q^n$ and $\tr \rho_{QR}^n$ are mapped to multi-replica correlators of the defect operator $\hat{\mathcal{E}}^{(n)}$, as in Eqs.~\eqref{eq:purities} and \eqref{eq:defect}. Prior to decoherence, the principal system $Q$ and the two-qubit reference system $R$ are initialized in the maximally entangled state
\begin{equation}
\label{eq:psi_QR}
    \ket{\Psi_{QR}} \equiv \frac{1}{\sqrt{4}} \sum_{\bh} \ket{\psi^{\bh}_h}_Q \otimes \ket{\bh}_R ,
\end{equation}
where $\qty{\ket{\psi^{\bh}_h}}$ are logical-$X$ eigenstates, as described in the main text. We then subject each link $\ell$ in the principal system to a bit-flip decoherence channel of the form $\mathcal{E}_{\ell}(\rho) \equiv (1 - p) \rho + p X_{\ell} \rho X_{\ell}$. The resulting density matrix on $QR$ is given by
\begin{equation}
    \begin{split}
        \rho_{QR} &\equiv \mathcal{E}(\dyad{\Psi_{QR}}) \\
        &= \frac{1}{4} \sum_{\bh, \bh'} \mathcal{E}(\dyad*{\psi^{\bh}_{h}}{\psi^{\bh'}_{h}})_{Q} \otimes \dyad{\bh}{\bh'}_{R} ,
    \end{split}
\end{equation}
where $\mathcal{E} \equiv \prod_{\ell} \mathcal{E}_{\ell}$ is the total channel. The corresponding density matrix on $Q$ alone is
\begin{equation}
    \rho_{Q} \equiv \tr_{R} \rho_{QR} = \frac{1}{4} \sum_{\bh} \mathcal{E}(\dyad*{\psi^{\bh}_{h}}{\psi^{\bh}_{h}}) .
\end{equation}

From these, we compute the purities $\tr \rho_{QR}^{n}$ and $\tr \rho_{Q}^{n}$. Note that the channel $\mathcal{E}$ is \emph{strongly symmetric} with respect to the logical-$X$ operators $\overline{X}_{\mu}$, and therefore $\mathcal{E}(\dyad{\psi^{\bh_{1}}_{h}}{\psi^{\bh_{2}}_{h}}) \, \mathcal{E}(\dyad{\psi^{\bh_{3}}_{h}}{\psi^{\bh_{4}}_{h}}) \propto \delta_{\bh_{2} \bh_{3}}$. From this, we obtain the expressions
\begin{widetext}
    \begin{subequations}
    \begin{align}
        \tr \rho_{QR}^{n} &= \frac{1}{4^{n}} \sum_{\bh_{1} \ldots \bh_{n}} \tr \Big[ \mathcal{E}(\dyad{\psi^{\bh_{1}}_{h}}{\psi^{\bh_{2}}_{h}}) \ldots \mathcal{E}(\dyad{\psi^{\bh_{n}}_{h}}{\psi^{\bh_{1}}_{h}}) \Big] , \label{eq:purity_QR} \\
        \tr \rho_{Q}^{n} &= \frac{1}{4^{n}} \sum_{\bh} \tr \Big[ \mathcal{E}(\dyad*{\psi^{\bh}_{h}}) \ldots \mathcal{E}(\dyad*{\psi^{\bh}_{h}}) \Big] . \label{eq:purity_Q} 
    \end{align}
\end{subequations}
Note that there are $n$ independent sums over ground states in Eq.~\eqref{eq:purity_QR}, but only a single sum in Eq.~\eqref{eq:purity_Q}. This leads to the two distinct boundary conditions discussed in the main text.

To write both of these expressions in a more suggestive form, let us express the total channel $\mathcal{E}$ in terms of Kraus operators as $\mathcal{E}(\rho) = \sum_{\be} p_{\be} X_{\be} \rho X_{\be}$, where $\be \equiv \qty{e_{\ell}}$ with $e_{\ell} = 0, 1$ denotes a collection of bit-flip errors, $X_{\be} \equiv \prod_{\ell} X_{\ell}^{e_{\ell}}$ denotes the full bit-flip error, and $p_{\be} \equiv (1 - p)^{2L^{2} - \abs{\be}} p^{\abs{\be}}$ is the corresponding probability. We then have
\begin{equation}
    \begin{split}
        \tr \Big[ \mathcal{E}(\dyad{\psi^{\bh_{1}}_{h}}{\psi^{\bh_{2}}_{h}}) \ldots \mathcal{E}(\dyad{\psi^{\bh_{n}}_{h}}{\psi^{\bh_{1}}_{h}}) \Big] &= \sum_{\be_{1} \ldots \be_{n}} p_{\be_{1}} \ldots p_{\be_{n}} \bra{\psi^{\bh_{1}}_{h}} X_{\be_{1}} X_{\be_{2}} \ket{\psi^{\bh_{1}}_{h}} \ldots \bra{\psi^{\bh_{n}}_{h}} X_{\be_{n}} X_{\be_{1}} \ket{\psi^{\bh_{n}}_{h}} \\
        &= \bra{\psi^{\bh_{1} \ldots \bh_{n}}_{h}} \hat{\mathcal{E}}^{(n)} \ket{\psi^{\bh_{1} \ldots \bh_{n}}_{h}} ,
    \end{split}
\end{equation}
\end{widetext}
where $\ket{\psi^{\bh_{1} \ldots \bh_{n}}_{h}} \equiv \ket{\psi^{\bh_{1}}_{h}} \otimes \ldots \ket{\psi^{\bh_{n}}_{h}}$, and $\hat{\mathcal{E}}^{(n)}$ is given by
\begin{equation}
    \begin{split}
        \hat{\mathcal{E}}^{(n)} &\equiv \sum_{\be_{1} \ldots \be_{n}} p_{\be_{1}} \ldots p_{\be_{n}} (X_{\be_{1}} X_{\be_{2}}) \otimes \ldots \otimes (X_{\be_{n}} X_{\be_{1}}) \\
        &= \prod_{\ell} \prod_{\alpha = 1}^{n} \qty[ (1 - p) + p X_{\ell}^{(\alpha)} X_{\ell}^{(\alpha + 1)} ] \\
        &= \prod_{\ell} \prod_{\alpha = 1}^{n} \exp{\gamma [X_{\ell}^{(\alpha)} X_{\ell}^{(\alpha + 1)} - 1]} ,
    \end{split} 
\end{equation}
with $\gamma$ defined by the relation $\tanh \gamma = p/(1-p)$. We thus arrive at Eq.~\eqref{eq:defect} of the main text.

\emph{Appendix B: Effective field theory.}--- While the decohered TFTC admits a particularly simple microscopic analysis, the general conclusions of this letter hold for a broad class of near-critical topological orders under decoherence. For example, similar microscopic statistical physics mappings can be carried out for $\mathbb{Z}_q$ TFTCs under generalized Pauli decoherence (see the Supplemental Material \cite{som} for details), as well as for certain tractable Hamiltonian deformations of non-abelian topological orders \cite{burnell_2012}. Here we will take a broader perspective by describing how generic near-critical topological codes under Pauli decoherence can be analyzed using an effective field theory viewpoint. For simplicity we focus on $\mathbb{Z}_2$ topological orders, although the present discussion can be immediately generalized.

In general Hamiltonian deformations of the toric code, such as $H' = H - \lambda \sum_{\ell} Z_{\ell} + \ldots$, the transition between topological and trivial phases is generically driven by the condensation of either the $e$ or $m$ anyon, while the other anyon remains gapped \cite{fn3} across the transition. Such a transition is once again generically described by the 3$d$ Ising* CFT \cite{fradkinPhaseDiagramsLattice1979}, where the Ising order parameter $\phi$ roughly represents the condensing anyon. The bulk action \eqref{eq:S_bulk} is therefore appropriate in the vicinity of a general topological-to-trivial transition with a single species of anyon condensing.

To consider more general decoherence channels beyond bit-flip decoherence, we construct the most general decoherence-induced perturbation to the Ising* CFT allowed by symmetries. For a completely general decoherence channel, the $n$th purities are again given by Eq.~\eqref{eq:purities}, with a modified defect operator $\hat{\mathcal{E}}^{(n)}$. Since this operator is approximately unity in the limit of weak decoherence, we can generally regard it as inserting a defect-like perturbation $\delta \mathcal{S}$ to the Ising* CFT along the $\tau = 0$ surface. This perturbation is constrained to satisfy the following symmetry requirements:
\begin{enumerate}
    \item Euclidean invariance and locality: assuming local i.i.d. decoherence, $\delta \mathcal{S} = \int \dd[2]{x} \mathcal{L}(\qty{\phi, \varepsilon})$ should be a local functional of the scaling fields $\phi^{(\alpha)}$ and $\varepsilon^{(\alpha)}$.
    \item Ising symmetry: since the Ising* CFT does not contain any $\mathbb{Z}_{2}$-odd fields (i.e., a single anyon cannot be created by any local operator), $\mathcal{L}$ must be symmetric under each replica's Ising symmetry $\phi^{(\alpha)} \mapsto -\phi^{(\alpha)}$. The most relevant terms will be constructed from combinations of the energy operator $\varepsilon^{(\alpha)}$, which is even under Ising symmetry.
    \item Replica symmetry: since a general defect operator $\hat{\mathcal{E}}^{(n)}$ couples replicas in a cyclic fashion, $\mathcal{L}$ must be invariant under the cyclic permutation $\phi^{(\alpha)} \mapsto \phi^{(\alpha + 1)}$, $\varepsilon^{(\alpha)} \mapsto \varepsilon^{(\alpha + 1)}$ (where $n + 1 \equiv 1$). 
    \item No single-replica terms: the set of all replica-symmetric perturbations can be classified as one-body (for example, $\sum_{\alpha} \varepsilon^{(\alpha)}$), two-body (for example, $\sum_{\alpha \alpha'} \varepsilon^{(\alpha)} \varepsilon^{(\alpha')}$), etc. However, since ``linear'' observables such as $\tr[O \rho_Q]$ cannot be affected whatsoever by channels acting outside the support of $O$, the replica limit of local single-replica observables must be identical to that of the unperturbed model. Consequently, one-body replica couplings cannot be renormalized by decoherence in the replica limit $n \to 1$ \cite{weinsteinNonlocalityEntanglementMeasured2023a,patil_highly_2024,nahum_bayesian_2025}. 
\end{enumerate}
Given these constraints, the most relevant terms which can be generated by decoherence are of the form
\begin{equation}
        \delta \mathcal{S} = \int \dd[2]{x} \sum_{\alpha \neq \alpha'} C_{\alpha \alpha'} \varepsilon^{(\alpha)}(0, x) \varepsilon^{(\alpha + 1)}(0, x) + \ldots ,
\end{equation}
where the constants $C_{\alpha \alpha'}$ depend only on $(\alpha - \alpha') \mod n$, and the ellipsis again denotes less relevant terms. We immediately see that $\delta \mathcal{S}$ can only contain perturbatively irrelevant terms, and the discussion of the main text carries through identically: since $r$ cannot be renormalized by the surface coupling $\delta \mathcal{S}$, a finite value of the couplings $C_{\alpha \alpha'}$ (and thereby a finite value of the microscopic decoherence strengths) is necessary to induce a transition in the replica limit.

An immediate application is to the Fradkin-Shenker model \cite{fradkinPhaseDiagramsLattice1979,tupitsyn_topological_2010,wu_phase_2012,verresenHiggsCondensatesAre2022}, described by the Hamiltonian
\begin{equation}
    H_{\text{FS}} \equiv - \sum_v A_v - \sum_p B_p - h \sum_{\ell} X_{\ell} - \lambda \sum_{\ell} Z_{\ell} .
\end{equation}
Whereas the transverse field $h$ provides quantum fluctuations to the $m$ anyons, the longitudinal field $\lambda$ provides quantum fluctuations to the $e$ anyons.

The $(h, \lambda)$ phase diagram of this model is well-known~\cite{fradkinPhaseDiagramsLattice1979}; in particular, although a nonzero longitudinal field $\lambda$ breaks the microscopic one-form symmetry of the TFTC, the 3$d$ Ising* transition at $(h, \lambda) = (h_c, 0)$ nevertheless extends into the $\lambda > 0$ phase diagram. The preceding discussion implies that the nonzero error threshold to local Pauli decoherence extends to nonzero values of $\lambda$, so long as the $e$ anyons remain gapped across the transition while $m$ anyons condense.

\makeatletter 

\let\addcontentsline\old@addcontentsline

\renewcommand\onecolumngrid{
\do@columngrid{one}{\@ne}%
\def\set@footnotewidth{\onecolumngrid}
\def\footnoterule{\kern-6pt\hrule width 1.5in\kern6pt}%
}

\renewcommand\twocolumngrid{
        \def\footnoterule{
        \dimen@\skip\footins\divide\dimen@\thr@@
        \kern-\dimen@\hrule width.5in\kern\dimen@}
        \do@columngrid{mlt}{\tw@}
}%

\newcommand{\thetitle}{\@title}

\makeatother

\onecolumngrid
\newpage

\renewcommand{\thefigure}{S\arabic{figure}}
\renewcommand{\theequation}{S\arabic{equation}}
\renewcommand{\thetable}{S\Roman{table}}
\renewcommand{\thesection}{S\Roman{section}}

\setcounter{secnumdepth}{2}
\setcounter{equation}{0}
\setcounter{section}{0}
\setcounter{figure}{0}
\setcounter{table}{0}
\setcounter{page}{1}

\newcommand{\LGT}{\text{LGT}}
\newcommand{\TFI}{\text{TFI}}
\newcommand{\vs}{\boldsymbol{s}}
\newcommand{\vh}{\boldsymbol{h}}

\renewcommand*{\thefootnote}{\fnsymbol{footnote}}
\setcounter{footnote}{0}
\footnotetext{%
\hypertarget{email}{}
\noindent \href{mailto:zackmw@caltech.edu}{zackmw@caltech.edu}
}
\setcounter{footnote}{0}
\renewcommand*{\thefootnote}{\arabic{footnote}}

\begin{center}
\textbf{\large Supplemental Material For: \thetitle}

\vspace{5mm}
Zack~Weinstein\textsuperscript{1,3,\hyperlink{email}{\textasteriskcentered}} and Samuel J. Garratt\textsuperscript{2,3} 

\vspace{1mm}

\textsuperscript{1}\textit{\small Department of Physics and Institute for Quantum Information and Matter, \\
\vspace{-0.5mm}
California Institute of Technology, Pasadena, California 91125, USA} \\
\vspace{-0.5mm}
\textsuperscript{2}\textit{\small Department of Physics, Princeton University, Princeton, New Jersey 08544, USA} \\
\vspace{-0.5mm}
\textsuperscript{3}\textit{\small Department of Physics, University of California, Berkeley, California 94720, USA} \\
\vspace{-0.5mm}
{\small (Dated: \today)}
\end{center}

\tableofcontents

\section{Background on the Transverse-Field Toric Code}
\label{app:tftc}
In this Appendix, we briefly review the physics of the transverse-field toric code (TFTC) studied in the main text. We place a particular emphasis on global aspects of Wegner duality which are less commonly discussed in the literature, but are crucial to the decoding transition studied in the present work.

\subsection{Model}
As discussed in the main text, we study the TFTC on an $L \times L$ square lattice on the torus, i.e., with periodic boundary conditions in each direction. Each link $\ell$ of the lattice is equipped with a qubit, on which the Pauli operators $X_{\ell}$ and $Z_{\ell}$ act. The Hamiltonian is given by
\begin{equation}
    H \equiv - \sum_{v} A_{v} - \sum_{p} B_{p} - h \sum_{\ell} X_{\ell} ,
\end{equation}
where $A_{v} \equiv \prod_{\ell \ni v} X_{\ell}$ and $B_{p} \equiv \prod_{\ell \in p} Z_{\ell}$ are the standard toric code vertex and plaquette stabilizers, respectively. 

The stabilizer toric code at $h = 0$ is especially well-known \cite{kitaev_fault-tolerant_2003}: since the stabilizers $A_{v}, B_{p}$ all mutually commute, the model is exactly solvable and the ground states $\ket{\psi_{0}}$ satisfy $A_{v} \ket{\psi_{0}} = B_{p} \ket{\psi_{0}} = + \ket{\psi_{0}}$. The ground-state manifold $\mathcal{C}_{0}$ is fourfold degenerate and its states are topologically ordered, allowing for a robust encoding of two logical qubits. Excitations with $A_{v} = -1$ and $B_{p} = -1$ are the $e$ and $m$ anyons respectively, which are self-bosons and mutual semions. The logical operators of the code space $\mathcal{C}_{0}$ coincide with two microscopic $\mathbb{Z}_{2}$ one-form symmetries \cite{wen_emergent_2019,mcgreevy_generalized_2023}, given by products of Pauli-$Z$ (Pauli-$X$) operators around non-contractible cycles of the torus:
\begin{equation}
    \overline{Z}_{\mu} \equiv \prod_{\ell \in C_{\mu}} Z_{\ell}, \quad \overline{X}_{\mu} \equiv \prod_{\ell \in \tilde{C}_{\mu}} X_{\ell}, \quad \mu = x, y ,
\end{equation}
where $C_{\mu}$ ($\tilde{C}_{\mu}$) is a non-contractible closed loop through the direct (dual) lattice about the $\mu$th cycle of the torus, with $\mu = x, y$. 

For $0 < h < h_{c} \approx 0.328$ \cite{wu_phase_2012}, topological order is preserved in the ground-state manifold of $H$. Specifically, $H$ exhibits an energetic splitting in its four low-lying states $\mathcal{C}_{h}$ which is exponentially small in $L$, separated from the remaining spectrum by an order-one gap. These ground states\footnote{We shall refer to all four nearly-degenerate low-lying states of $H$ simply as ``ground states'', since they become degenerate in the thermodynamic limit.} remain topologically ordered and locally indistinguishable up to exponentially small corrections in $L$ \cite{hastings_quasiadiabatic_2005}. It will be useful to note that the logical-$X$ operators $\overline{X}_{\mu}$ remain an exact one-form symmetry of $H$ for all values of $h$, and so the four ground states of $H$ can be unambiguously chosen as logical-$X$ eigenstates. We denote these states as $\qty{\ket{\psi^{\bh}_{h}}}$, where $\bh = (\eta_{x}, \eta_{y})$ with $\eta_{x, y} = \pm 1$ denotes the eigenvalues of the logical-$X$ operators:
\begin{equation}
    \overline{X}_{\mu} \ket{\psi^{\bh}_{h}} = \eta_{\mu} \ket{\psi^{\bh}_{h}}, \quad \mu = x, y .
\end{equation}
In each state $\ket{\psi^{\bh}_{h}}$, a phase of $\eta_{\mu}$ is acquired when a bare $m$ anyon is transported around the $\mu$th cycle of the torus. While the microscopic logical-$Z$ operators no longer commute with $H$ for $h > 0$, it is possible to construct ``fattened'' versions of the logical-$Z$ operators for all $h < h_{c}$ via quasiadiabatic continuation \cite{hastings_quasiadiabatic_2005}.

Physically, the effect of nonzero $h$ is to provide dynamics and quantum fluctuations to the $m$ anyons, which remain gapped throughout the topological phase $h < h_{c}$. As $h \to h_{c}$ from below, the $m$ anyons condense and destroy the topological order. The physics of the single topologically trivial ground state at large $h$ can be understood by regarding the excitations $X_{\ell} = -1$ as segments of $\mathbb{Z}_{2}$ electric field lines, which form closed loops through the dual lattice due to the ``Gauss's law'' constraint $A_{v} = +1$. These electric field lines exhibit a finite line tension for $h > h_{c}$, and so creating a pair of $e$ anyons costs an energy which is linear in their separation. The $h > h_{c}$ phase is therefore a \emph{confined} phase. As $h \to h_{c}$ from above, electric field lines lose their line tension and condense. In modern parlance, this is a 1-form spontaneous symmetry breaking transition \cite{mcgreevy_generalized_2023,pace_exact_2023}.

\subsection{Wegner Duality}
The quantum phase transition in the TFTC at $h_{c}$ is described by the (2+1)$d$ Ising* universality class. To see this, we first constrain ourselves to the sector $A_{v} = +1$ of Hilbert space, which contains the ground-state manifold. We can then trivialize this constraint by defining a new set of Pauli operators $\sigma^{x}_{p}, \sigma^{z}_{p}$ on at the centers of each plaquette (i.e., the sites of the dual lattice), and making the identification
\begin{equation}
    \label{eq:wegner_1}
    X_{\ell} \longleftrightarrow \sigma^{z}_{p} \sigma^{z}_{q} ,
\end{equation}
where $p, q$ are the two nearest-neighbor plaquettes bisected by the link $\ell$. This identification guarantees that $A_{v} \longleftrightarrow \prod_{\expval{pq} \in v} (\sigma^{z}_{p} \sigma^{z}_{q}) = 1$, where $v$ is regarded as a dual-lattice plaquette in the second expression. Since $X_{\ell}$ anticommutes with the two $B_{p}$ on the two plaquettes $p,q$ bisected by $\ell$, we also make the identification
\begin{equation}
    \label{eq:wegner_2}
    B_{p} \longleftrightarrow \sigma^{x}_{p} .
\end{equation}
Altogether, we find that the TFTC restricted to the sector $A_{v} = +1$ is dual to the (2+1)$d$ transverse-field Ising model (TFIM),
\begin{equation}
    H_{\TFI} \equiv - \sum_{p} \sigma^{x}_{p} - h \sum_{\expval{pq}} \sigma^{z}_{p} \sigma^{z}_{q} .
\end{equation}
Physically, this duality identifies the $m$ anyons $B_{p} = -1$ of the TFTC with the spin-flip excitations $\sigma^{x}_{p} = -1$ of the TFIM, and electric field lines $X_{\ell} = -1$ of the TFTC with the domain walls $\sigma^{z}_{p} \sigma^{z}_{q}$ of the TFIM. The topological (trivial) phase of the TFTC thus corresponds to paramagnetic (ferromagnetic) phase of the TFIM.

While this duality transformation correctly relates the bulk physics of these two models, the identifications \eqref{eq:wegner_1} and \eqref{eq:wegner_2} clearly cannot be globally exact. With some care, we can promote these identifications to \emph{exact} operator identities as follows:
\begin{itemize}
    \item The second expression \eqref{eq:wegner_2} is straightforward: since $\prod_{p} B_{p} = 1$, we must necessarily restrict to the parity-even sector $\prod_{p} \sigma^{x}_{p} = 1$ of the TFIM. In other words, there are no operators of the TFTC which are charged under the dual $\mathbb{Z}_{2}$ symmetry. For this reason, we refer to the critical point of the TFTC as the Ising* CFT, where the star indicates that $\mathbb{Z}_{2}$-charged scaling operators are not present in the CFT.
    
    \item The first expression \eqref{eq:wegner_1} requires a bit more care. Within each logical sector $\overline{X}_{\mu} = \eta_{\mu}$ ($\mu = x, y$), we have the identity
    \begin{equation}
        \prod_{\expval{pq} \in \tilde{C}_{\mu}} \sigma^{z}_{p} \sigma^{z}_{q} = \eta_{\mu} ,
    \end{equation}
    i.e., we should regard the TFIM as exhibiting periodic (antiperiodic) boundary conditions about the $\mu$th cycle of the torus whenever $\eta_{\mu} = +1$ ($\eta_{\mu} = -1$). We can make these boundary conditions explicit by modifying Eq.~\eqref{eq:wegner_1} to\footnote{Within each logical sector, we can think of $U^{\bh}_{pq}$ as a flat, non-dynamical $\mathbb{Z}_{2}$ gauge field, fixed to a particular gauge. This simply encodes the boundary conditions of the TFIM. In a more complete formulation of Wegner duality, the TFTC is dual to a TFIM coupled to a flat \emph{dynamical} $\mathbb{Z}_{2}$ gauge field. A simple method of arriving at this duality is to start from the self-duality of the Fradkin-Shenker model~\cite{fradkinPhaseDiagramsLattice1979}, and to take the Higgs coupling to zero.}
    \begin{equation}
        \label{eq:wegner_3}
        X_{\ell} \longleftrightarrow U_{pq}^{\bh} \sigma^{z}_{p} \sigma^{z}_{q} , \quad U^{\bh}_{pq} \equiv \begin{cases}
            +1, & \expval{pq} \not\in \Gamma_{x}, \Gamma_{y} \\
            \eta_{y}, & \expval{pq} \in \Gamma_{x} \\
            \eta_{x}, & \expval{pq} \in \Gamma_{y}
        \end{cases} ,
    \end{equation}
    where we've fixed two particular representative non-contractible loops $\Gamma_{x}, \Gamma_{y}$ through the direct lattice about the two cycles of the torus. Each logical operator $\overline{X}_{x}$ ($\overline{X}_{y}$) must necessarily cross $\Gamma_{y}$ ($\Gamma_{x}$) an odd number of times, ensuring that $\overline{X}_{\mu} = \eta_{\mu}$. 
\end{itemize}

In summary, the $A_{v} = 1$ sector of the TFTC is exactly dual to a TFIM on the dual lattice, so long as (a) the TFIM is restricted to its parity-even sector, and (b) the TFIM includes both periodic and antiperiodic boundary conditions in its Hilbert space. In the $h < h_{c}$ phase, the dual TFIM is paramagnetic, and the ground-state energetic splitting between these boundary conditions is exponentially small in $L$, leading to a four-dimensional ground-state manifold as required. In the $h > h_{c}$ phase, the dual TFIM is ferromagnetic and antiperiodic boundary conditions exhibit an $\mathcal{O}(L)$ ground-state energy cost above the $\bh = (+1, +1)$ sector. Combined with the requirement $\prod_{p} \sigma^{x}_{p} = +1$ which forbids symmetry-broken ground states, this leads to a unique gapped ground state as required.

\section{Details on the Statistical Physics Mapping}
\label{app:details}
In this Appendix, we provide technical details for the statistical physics mapping described in the main text. As discussed in the main text, both $n$th purities $\tr \rho_{QR}^n$ and $\tr \rho_Q^n$ can be simply expressed in terms of matrix elements of the form
\begin{equation}
    \mathcal{Z}^{\bh_1 \ldots \bh_n} \equiv \bra{\psi^{\bh_1 \ldots \bh_n}_h} \hat{\mathcal{E}}^{(n)} \ket{\psi^{\bh_1 \ldots \bh_n}_h} ,
\end{equation}
where $\ket{\psi^{\bh_1 \ldots \bh_n}_h} \equiv \ket{\psi^{\bh_1}_h} \otimes \ldots \ket{\psi^{\bh_n}_h}$ is a tensor product of logical-$X$ eigenstates $\ket{\psi^{\bh}_h}$, and the defect operator $\hat{\mathcal{E}}^{(n)}$ is given by Eq.~\eqref{eq:defect}.
To develop a statistical physics interpretation of these matrix elements, we utilize the quantum-classical mapping \cite{fradkin_order_1978,kogutIntroductionLatticeGauge1979,sachdevQuantumPhaseTransitions2011}. We first note that each of the basis states $\ket{\psi^{\bh}_{h}}$ can be conveniently represented via imaginary-time evolution from the corresponding stabilizer code state $\ket{\psi^{\bh}_{0}}$:
\begin{equation}
    \ket{\psi^{\bh}_{h}} = \lim_{\beta \to \infty} \frac{e^{- \beta H / 2} \ket{\psi^{\bh}_{0}}}{\sqrt{\bra{\psi^{\bh}_{0}} e^{-\beta H} \ket{\psi^{\bh}_{0}}}} .
\end{equation}
As $\beta \to \infty$, the Gibbs factor $e^{-\beta H / 2}$ becomes proportional to a projector onto the ground-state manifold of $H$. Furthermore, since $H$ commutes with the logical-$X$ operators, the eigenvalues $\bh$ of the resulting state are preserved under imaginary-time evolution. While this equation is exact in the strict limit $\beta \to \infty$, since $H$ is gapped, it also holds up to exponentially small corrections in $L$ when $\beta \sim \text{poly}(L)$.

To derive a quantum-classical mapping for the states $\ket{\psi^{\bh}_{h}}$, it is useful to first perform a duality transformation. As discussed in Appendix~\ref{app:tftc}, the TFTC is Wegner-dual to the (2+1)$d$ transverse-field Ising model (TFIM). Specifically, in each logical-$X$ sector $\overline{X}_{\mu} = \eta_{\mu}$, we have the identity
\begin{equation}
    \bra{\psi^{\bh}_{0}} e^{-\beta H} \ket{\psi^{\bh}_{0}} = \bra{+} e^{-\beta H_{\TFI}^{\bh}} \ket{+}, \quad H_{\TFI}^{\bh} \equiv - \sum_{p} \sigma^{x}_{p} - h \sum_{\expval{pq}} U^{\bh}_{pq} \sigma^{z}_{p} \sigma^{z}_{q} ,
\end{equation}
where $\ket{+}$ is the simultaneous $+1$ eigenstate of each $\sigma^{x}_{p}$, and $U^{\bh}_{pq}$ is defined as in Eq.~\eqref{eq:wegner_3}. We note that this equality is exact. 

The quantum-classical mapping for $\bra{+} e^{-\beta H_{\TFI}^{\bh}} \ket{+}$ now proceeds in standard fashion \cite{sachdevQuantumPhaseTransitions2011}: dividing $\beta = \frac{\varepsilon}{M}$ into $M$ imaginary time slices, and inserting $M - 1$ resolutions of the identity in the Pauli-$Z$ basis, we obtain
\begin{equation}
    \bra{+} e^{-\beta H_{\TFI}^{\bh}} \ket{+} = \frac{1}{2^{L^{2}}} \sum_{\boldsymbol{\sigma}_{0} \ldots \boldsymbol{\sigma}_{M}} \bra{\boldsymbol{\sigma}_{M}} e^{- \varepsilon H^{\bh}_{\TFI}} \ket{\boldsymbol{\sigma}_{M-1}} \ldots \bra{\boldsymbol{\sigma}_{1}} e^{-\varepsilon H^{\bh}_{\TFI}} \ket{\boldsymbol{\sigma}_{0}} ,
\end{equation}
where $\boldsymbol{\sigma}_{m} \equiv \qty{\sigma_{p, m}}$ is the full state of the system along the $m$th time slice, and we have written $\ket{+} = 2^{-L^{2}/2} \sum_{\boldsymbol{\sigma}} \ket{\boldsymbol{\sigma}}$. Assuming $\varepsilon$ is sufficiently small, we can evaluate each matrix element by separating the Pauli-$X$ and Pauli-$Z$ terms:
\begin{equation}
    \begin{split}
        \bra{\boldsymbol{\sigma}_{m}} e^{- \varepsilon H^{\bh}_{\TFI}} \ket{\boldsymbol{\sigma}_{m-1}} &\approx \bra{\boldsymbol{\sigma}_{m}} e^{- \varepsilon H^{\bh}_{\TFI}} \ket{\boldsymbol{\sigma}_{m-1}} \\
        &= e^{\varepsilon h \sum_{\expval{pq}} U_{pq}^{\bh} \sigma_{p,m} \sigma_{q, m} } \prod_{p} \bra{\sigma_{p, m}} e^{\varepsilon \sigma^{x}_{p}} \ket{\sigma_{p, m-1}} \\
        &= e^{\sum_{\expval{pq}} J^{\bh}_{pq} \sigma_{p,m} \sigma_{q, m} } \prod_{p} \qty[ \frac{\cosh \varepsilon}{e^{J^{\parallel}}} e^{J^{\parallel} \sigma_{p, m} \sigma_{p, m-1}} ] ,
    \end{split}
\end{equation}
where we've defined the in-plane couplings $J^{\bh}_{pq} \equiv \varepsilon h U^{\bh}_{pq}$ and the inter-plane coupling $J^{\parallel} = - \frac{1}{2} \log \tanh \varepsilon \approx \frac{1}{2} \log \frac{1}{\varepsilon}$. Altogether, we obtain the result
\begin{subequations}
    \begin{align}
        \bra{\psi^{\bh}_{0}} e^{-\beta H} \ket{\psi^{\bh}_{0}} &\propto \sum_{\qty{\sigma}} e^{-S_{\Ising}^{\bh}[\sigma]} , \\
        S_{\Ising}^{\bh}[\sigma] &= - \sum_{m = 1}^{M} \sum_{\expval{pq}} J^{\bh}_{pq} \sigma_{p, m} \sigma_{q, m} - \sum_{m = 1}^{M} \sum_{p} J^{\parallel} \sigma_{p, m} \sigma_{p, m - 1} \\
        &= - \sum_{\expval{ij}} J^{\bh}_{ij} \sigma_{i} \sigma_{j} , \notag
    \end{align}
\end{subequations}
where in the final line, we have defined the spacetime lattice index $i = (p, m)$, as well as the spacetime couplings $J^{\bh}_{ij}$. We see that $\bra{\psi^{\bh}_{0}} e^{-\beta H} \ket{\psi^{\bh}_{0}}$ is proportional to the partition function of an extremely anisotropic 3$d$ Ising model, with (anti)periodic boundary conditions in the $x$ and $y$ directions, and open boundary conditions at imaginary times $\tau = \pm \frac{\beta}{2}$ (corresponding to $m = 0, M$).

The quantum-classical mapping for $\mathcal{Z}^{\bh_{1} \ldots \bh_{n}}$ proceeds analogously: besides a single imaginary time slice at $\tau = 0$ (i.e., $m = M/2$), the mapping proceeds identically as for $n$ decoupled TFTCs. At $\tau = 0$, we have an insertion of the defect operator $\hat{\mathcal{E}}^{(n)}$, where $X_{\ell}^{(\alpha)}$ is replaced by $(\sigma^{z}_{p} U^{\bh}_{pq} \sigma^{z}_{q})^{(\alpha)}$. We thus obtain
\begin{equation}
    \begin{split}
        \mathcal{Z}^{\bh_{1} \ldots \bh_{n}} &\propto \sum_{\qty{\sigma}} \exp \qty{ - \sum_{\alpha = 1}^{n} S_{\Ising}^{\bh}[\sigma^{(\alpha)}] } \bra{\boldsymbol{\sigma}^{(1)}_{\tau = 0} \ldots \boldsymbol{\sigma}^{(n)}_{\tau = 0}} \hat{\mathcal{E}}^{(n)} \ket{\boldsymbol{\sigma}^{(1)}_{\tau = 0} \ldots \boldsymbol{\sigma}^{(n)}_{\tau = 0}} \\
        &\propto \sum_{\qty{\sigma}} \exp \qty{ - \sum_{\alpha = 1}^{n} S_{\Ising}^{\bh}[\sigma^{(\alpha)}] - \delta S[\sigma] } ,
    \end{split}
\end{equation}
with $\delta S[\sigma]$ given by Eq.~\eqref{eq:defect_statmech} of the main text.

\section{\texorpdfstring{$\mathbb{Z}_q$}{Zq} Transverse-Field Toric Code Under Shift Decoherence}
As discussed in the main text, our results can be generalized beyond the TFTC to a broad family of topological-to-trivial quantum phase transitions driven by a condensing self-boson. As a simple and concrete illustration, here we briefly discuss the $\mathbb{Z}_q$ TFTC under ``shift'' decoherence.

As in the $\mathbb{Z}_2$ case, we consider an $L \times L$ square lattice with periodic boundary conditions. Each link $\ell$ of the lattice hosts a $q$-state qudit with the local Hilbert space $\mathcal{H}^{(q)} \equiv \text{span} \qty{\ket{0}, \ket{1}, \ldots , \ket{q - 1}}$. We define the clock operator $Z$ and the shift operator $X$ on this qudit space to act as follows:
\begin{equation}
    Z \ket{k} = \omega^k \ket{k}, \quad X \ket{k} = \ket{k + 1 \mod q} ,
\end{equation}
where $\omega \equiv e^{2\pi i / q}$. 

The unperturbed $\mathbb{Z}_q$ toric code is defined by the Hamiltonian
\begin{equation}
    H_0^{(p)} \equiv - \frac{1}{2} \sum_v (A_v + A_v^{\dag}) - \frac{1}{2} \sum_p (B_p + B_p^{\dag}) ,
\end{equation}
where $A_v$ and $B_p$ are respectively defined as follows:
\begin{equation}
    A_{v} \equiv \begin{tikzpicture}[baseline={(current bounding box.center)}]
        \draw[line width = 1.0pt, gray!50] (1, 0) -- (-1, 0);
        \draw[line width = 1.0pt, gray!50] (0, 1) -- (0, -1);
        \node at (0.5, 0) {\large $X$};
        \node at (-0.5, 0) {\large $X^{\dag}$};
        \node at (0, 0.5) {\large $X$};
        \node at (0, -0.5) {\large $X^{\dag}$};
        \node at (0, 0) {\large $v$}; 
    \end{tikzpicture}, \quad \quad B_{p} \equiv \begin{tikzpicture}[baseline={(current bounding box.center)}]
        \draw[line width = 1.0pt, gray!50] (0, 0) -- (0, 1.5) -- (1.5, 1.5) -- (1.5, 0) -- cycle;
        \node at (0, 0.75) {\large $Z^{\dag}$};
        \node at (0.75, 0) {\large $Z$};
        \node at (1.5, 0.75) {\large $Z$};
        \node at (0.75, 1.5) {\large $Z^{\dag}$};
        \node at (0.75, 0.75) {\large $p$};
    \end{tikzpicture} .
\end{equation}
It is readily verified that each pair of $A_v$ and $B_p$ operators commute, and so the unperturbed ground state is defined by $A_v = B_p = +1$. Excitations with $A_v = \omega^k$ and $B_p = \omega^k$ are the $e^k$ and $m^k$ anyons, respectively.

Similar to the $\mathbb{Z}_2$ case, the $\mathbb{Z}_q$ TFTC defined by the Hamiltonian
\begin{equation}
    H^{(q)} \equiv H^{(q)}_0 - \frac{h}{2} \sum_{\ell} (X_{\ell} + X_{\ell}^{\dag}) ,
\end{equation}
where the effect of the transverse field $h$ is to provide quantum fluctuations to the $m$ anyons. This simple deformation of the $\mathbb{Z}_q$ toric code allows for a direct microscopic derivation of a statistical physics mapping for the R\'enyi coherent information. Nevertheless, as discussed in Appendix B of the End Matter, we can also consider more general Hamiltonians which provide nontrivial quantum fluctuations to the $e$ anyons, as long as they remain gapped as the $m$ anyons condense.

Just as the $\mathbb{Z}_2$ TFTC is dual to the $(2+1)d$ transverse-field Ising model (see Appendix~\ref{app:tftc}), the $\mathbb{Z}_q$ TFTC is dual to the $(2+1)d$ transverse-field clock model \cite{horn_hamiltonian_1979,cobanera_bond-algebraic_2011}, described by the Hamiltonian
\begin{equation}
    \tilde{H}^{(q)} = - \frac{1}{2} \sum_{p} (\tilde{X}_p + \tilde{X}_p^{\dag}) - \frac{h}{2} \sum_{\expval{p p'}} (\tilde{Z}^{\dag}_p \tilde{Z}_{p'} + \tilde{Z}_p \tilde{Z}_p^{\dag}) ,
\end{equation}
where $\tilde{Z}_p$ and $\tilde{X}_p$ are a new set of clock and shift operators defined on the dual lattice sites (i.e., the plaquettes) $p$. The latter model undergoes a continuous quantum phase transition in the 3$d$ XY universality class (i.e., the 3$d$ O(2) Wilson-Fisher CFT) for\footnote{For $q = 3$, the transition in both models is first-order \cite{janke_threedimensional_1997}. Although one might conjecture that the absence of a diverging correlation length near the transition will only help in establishing a finite error threshold, we cannot make any universal statements in this setting.} $q \geq 4$ \cite{oshikawa_ordered_2000,hove_criticality_2003,lou_emergence_2007}. Consequently, the $\mathbb{Z}_q$ TFTC exhibits a quantum phase transition in the 3$d$ XY* universality class for $q \geq 4$, where the star indicates that U(1)-charged scaling operators are absent from the CFT. For $h < h_c^{(q)}$ the $\mathbb{Z}_q$ TFTC exhibits a $q^2$-fold nearly degenerate ground-state manifold, which can be used to encode two logical $q$-state qudits. For $h > h_c^{(q)}$ the $m$ anyons are condensed, and the model exhibits a unique gapped ground state.

As in the main text, we are interested in how robustly quantum information is protected within the ground-state manifold of $H^{(q)}$. Towards this end, we consider ``shift'' decoherence acting on each link $\ell$, described by the channel
\begin{equation}
    \mathcal{E}_{\ell}(\rho) = (1 - p) \rho + \sum_{k = 1}^{q - 1} p_k X_{\ell}^k \rho X_{\ell}^{\dag k} ,
\end{equation}
such that $\sum_{k = 1}^{q - 1} p_k = p$. The precise choice of numbers $p_k$ will be unimportant for our purposes. This problem can be analyzed identically to the $\mathbb{Z}_2$ case in the main text: starting from a maximally entangled state $\ket{\psi_{QR}}$ between the principal system $Q$ and a two-qudit reference system $R$, we use the dual clock model representation and the quantum-classical mapping to represent the purities $\tr \rho_{QR}^n$ and $\tr \rho_Q^n$ as partition functions of an $n$-fold replicated 3$d$ classical clock model. The decoherence functions as an inter-replica defect along the $\tau = 0$ slice, which couples the clock models by their energy densities $\tilde{Z}_p^{\dag} \tilde{Z}_{p'} + \tilde{Z}_p \tilde{Z}_{p'}^{\dag}$ in a cyclic fashion. In the vicinity of the phase transition at $h_c^{(q)}$, we can describe each replica by an XY* CFT with a complex order parameter $\psi^{(\alpha)}(\tau, x)$ ($\alpha = 1, \ldots , n$), and so both purities are described by an action of the form
\begin{subequations}
    \begin{align}
        \mathcal{S}[\qty{\psi}] &= \sum_{\alpha = 1}^n \mathcal{S}_0[\psi^{(\alpha)}] + \delta \mathcal{S}[\qty{\psi}] + \ldots , \\
        \mathcal{S}_0[\psi^{(\alpha)}] &\equiv \mathcal{S}_{\text{XY}}[\psi^{(\alpha)}] + r \int \dd{\tau} \dd[2]{x} \varepsilon^{(\alpha)}(\tau, x) , \\
        \delta \mathcal{S}[\qty{\psi}] &\equiv - \tilde{\gamma} \int \dd[2]{x} \sum_{\alpha = 1}^n \varepsilon^{(\alpha)}(0, x) \varepsilon^{(\alpha + 1)}(0, x) . 
    \end{align}
\end{subequations}
Here $\mathcal{S}_{\text{XY}}[\psi^{(\alpha)}]$ formally represents the XY* CFT in replica $\alpha$, $\varepsilon^{(\alpha)}(\tau, x) \sim \abs{\psi^{(\alpha)}(\tau, x)}^2$ is the energy operator in replica $\alpha$, and the ellipsis denotes less relevant terms. These less relevant terms include clock anisotropies such as $[\psi^{(\alpha)}]^q + \text{c.c.}$, which are irrelevant to the Wilson-Fisher critical point for $q \geq 4$ \cite{oshikawa_ordered_2000}. As in the main text, the perturbation $r \propto h_{c}^{(q)} - h$ pushes the system into its paramagnetic phase, and $\tilde{\gamma}$ is proportional to $p$ for sufficiently weak measurement strengths. The $n$th R\'enyi coherent information is once again proportional to the free energy cost of locking the boundary conditions of the $n$ replicas together.

For the XY CFT, the scaling dimension of the energy operator is $\Delta_{\varepsilon} \approx 1.5$ \cite{campostrini_critical_2001}, and so $\delta \mathcal{S}$ is again an irrelevant perturbation to the critical point. Moreover, since it is a surface coupling, it cannot renormalize the bulk temperature $r$. Therefore, by an identical argument to that of the main text, the $\mathbb{Z}_q$ TFTC exhibits a finite threshold to shift decoherence in the vicinity of the quantum critical point.

\section{Strong-To-Weak Spontaneous Symmetry Breaking in the Nearly Critical Quantum Ising Model}
\label{app:swssb}
In the main text, we studied the intrinsic error threshold of the TFTC subjected to bit-flip decoherence. A useful analytical handle was offered by the TFTC's Wegner duality to the (2+1)$d$ TFIM. Given that the paramagnetic TFIM cannot encode any quantum information, it is natural to ask whether the error threshold in the TFTC has any physical significance in the dual TFIM. Here we point out that the error threshold describes a \emph{strong-to-weak spontaneous symmetry breaking} (SWSSB) transition \cite{leeQuantumCriticalityDecoherence2023,lessa2024strongtoweakspontaneoussymmetrybreaking,sala2024spontaneousstrongsymmetrybreaking} in these dual variables. Consequently, the results of the main text can be reinterpreted as stating that a finite strength of strongly symmetric decoherence is necessary to induce SWSSB, even as the transition to the ferromagnetic phase is approached.

For concreteness, let us reframe the problem entirely starting from the TFIM. We also generalize from two spatial dimensions to $d$ spatial dimensions. The TFIM Hamiltonian is defined by
\begin{equation}
    H_{\TFI} \equiv - \sum_{\expval{v v'}} Z_{v} Z_{v'} - g \sum_{v} X_{v} ,
\end{equation}
where $Z_{v}, X_{v}$ are Pauli operators on the vertices of a $d$-dimensional cubic lattice. So long as $g > g_{c}$, the model is in its paramagnetic phase, and exhibits a unique ground state $\ket{\psi_{g}}$ which is invariant under the $\mathbb{Z}_{2}$ symmetry $\Pi \equiv \prod_{v} X_{v}$. To this state, we apply the strongly $\mathbb{Z}_{2}$-symmetric decoherence channel
\begin{equation}
    \label{eq:swssb_decoherence}
    \mathcal{E}_{v v'}(\rho) \equiv (1 - p) \rho + p Z_{v} Z_{v'} \rho Z_{v} Z_{v'}
\end{equation}
to each nearest-neighbor link $\expval{v v'}$. The resulting mixed state is $\rho_{g} \equiv \mathcal{E}(\dyad{\psi_{g}})$, where $\mathcal{E} \equiv \prod_{\expval{v v'}} \mathcal{E}_{v v'}$.

The mixed state $\rho_{g}$ is \emph{strongly symmetric}, i.e., $\Pi \rho_{g} = \rho_{g} \Pi = \rho_{g}$. However, at sufficiently large decoherence strengths $p$, we expect that this strong $\mathbb{Z}_{2}$ symmetry can be spontaneously broken down to a residual weak symmetry \cite{leeQuantumCriticalityDecoherence2023,lessa2024strongtoweakspontaneoussymmetrybreaking,sala2024spontaneousstrongsymmetrybreaking}. Here we define SWSSB by long-range order in the R\'enyi-1 correlator \cite{weinstein_efficient_2025},
\begin{equation}
    R_{1}(x, y) \equiv \tr[Z_{x} Z_{y} \sqrt{\rho_{g}} Z_{x} Z_{y} \sqrt{\rho_{g}}] ,
\end{equation}
in the absence of conventional long-range order in $\tr[Z_{x} Z_{y} \rho_{g}]$, where $x$ and $y$ are sites of the lattice. Using this observable, SWSSB is simply defined as a particular pattern of conventional spontaneous symmetry breaking in the canonical purification state $\lvert \sqrt{\rho_{g}} \rangle \! \rangle$ \cite{weinstein_efficient_2025}.

SWSSB in the decohered TFIM has been thoroughly studied in the limit $g \to \infty$ \cite{leeQuantumCriticalityDecoherence2023,lessa2024strongtoweakspontaneoussymmetrybreaking,sala2024spontaneousstrongsymmetrybreaking} (see also the Supplemental Material of Ref.~\cite{weinstein_efficient_2025}), where $\ket{\psi_{g = \infty}} = \ket{+}^{\otimes N}$ is a simple $\mathbb{Z}_{2}$-symmetric product state. In $d = 2$, this problem is Wegner-dual to the stabilizer toric code under bit-flip decoherence, and the SWSSB transition occurs at $p_{c} \simeq 0.109$ and is described by the Nishimori random-bond Ising model. An analogous transition occurs in $d \geq 2$, while in $d = 1$ there is no stable SWSSB phase (i.e., SWSSB occurs only at the maximal value $p = \frac{1}{2}$).

Here we are interested in what happens for finite $g$ larger than $g_{c}$, and in particular what happens to the critical SWSSB threshold $p_{c}(g)$ as $g \to g_{c}$ from above. Towards this end, we once again rely on the replica trick: we first introduce the R\'enyi-$n$ correlators,
\begin{equation}
    R_{n}(x, y) \equiv \frac{\tr[Z_{x} Z_{y} \rho_{g}^{n/2} Z_{x} Z_{y} \rho_{g}^{n/2}]}{\tr \rho_{g}^{n}} ,
\end{equation}
which is defined for all $n \geq 1$, but is conveniently evaluated along the even integers. In principle, the replica limit $n \to 1$ recovers $R_{1}(x, y)$. After identical manipulations to those of Appendix~\ref{app:details}, we arrive at the expressions
\begin{equation}
    \tr \rho_{g}^{n} = \bra{\psi_{g}^{\otimes n}} \hat{\mathcal{E}}^{(n)} \ket{\psi_{g}^{\otimes n}}, \quad \tr[Z_{x} Z_{y} \rho_{g}^{n/2} Z_{x} Z_{y} \rho_{g}^{n/2}] = \bra{\psi_{g}^{\otimes n}} (Z_{x} Z_{y})^{(1)} (Z_{x} Z_{y})^{(1 + \frac{n}{2})} \hat{\mathcal{E}}^{(n)} \ket{\psi_{g}^{\otimes n}} ,
\end{equation}
where $\hat{\mathcal{E}}^{(n)} \equiv \prod_{\expval{v v'}} \prod_{\alpha = 1}^{n} e^{\gamma[Z^{(\alpha)}_{v} Z^{(\alpha + 1)}_{v} - 1]}$, with $\gamma \equiv \tanh^{-1}[p/(1-p)]$ as in the main text. Employing the quantum classical mapping as before, we find that the $n$th purity $\tr \rho_{g}^{n}$ is proportional to the partition function of a ($d$+1)-dimensional replica statistical physics model,
\begin{subequations}
    \begin{align}
        \tr \rho_{g}^{n} \propto \mathcal{Z}^{(n)} &\equiv \sum_{\qty{\sigma}} \exp \qty{ - \sum_{\alpha = 1}^{n} S_{\Ising}[\sigma^{(\alpha)}] - \delta S[\sigma] } , \\
        S_{\Ising}[\sigma^{(\alpha)}] &\equiv - \sum_{\expval{ij}} J_{ij} \sigma^{(\alpha)}_{i} \sigma^{(\alpha)}_{j}, \\
        \delta S[\sigma] &\equiv - \gamma \sum_{\expval{ij} | \tau = 0} \sigma_{i}^{(\alpha)} \sigma_{j}^{(\alpha)} \sigma_{i}^{(\alpha + 1)} \sigma_{j}^{(\alpha + 1)} ,
    \end{align}
\end{subequations}
while $R_{n}(x, y)$ is given by a correlation function with respect to this replica model:
\begin{equation}
\label{eq:R_n}
    \begin{split}
        R_{n}(x, y) &= \expval{\sigma_{x}^{(1)} \sigma_{y}^{(1)} \sigma_{x}^{(1 + \frac{n}{2})} \sigma_{y}^{(1 + \frac{n}{2})} }_{\gamma}^{(n)} \\
        &= \frac{1}{\mathcal{Z}^{(n)}} \sum_{\qty{\sigma}} \sigma_{x}^{(1)} \sigma_{y}^{(1)} \sigma_{x}^{(1 + \frac{n}{2})} \sigma_{y}^{(1 + \frac{n}{2})} \exp \qty{ - \sum_{\alpha = 1}^{n} S_{\Ising}[\sigma^{(\alpha)}] - \delta S[\sigma] } .
    \end{split}
\end{equation}
In other words, $R_{n}(x, y)$ is a two-point correlation function of the order parameter $\sigma^{(\alpha)}_{i} \sigma^{(\alpha')}_{i}$, for $\alpha = 1$ and $\alpha' = 1 + \frac{n}{2}$.

In dimensions $d \geq 2$, the phase diagram of the models defined by the partition functions $\mathcal{Z}^{(n)}$ follows identically to the analysis of the main text. For $g > g_c$ and $\gamma < \gamma_c^{(n)}(g)$, the model is disordered; for $g > g_c$ and $\gamma > \gamma_c^{(n)}(g)$, the $\tau = 0$ surface exhibits long-range correlations in $\sigma_{i}^{(\alpha)} \sigma_{i}^{(\alpha ')}$ while the bulk remains disordered; and for $g < g_c$, the bulk and boundary both order. The R\'enyi-$n$ correlator becomes long-range ordered in the latter two phases. Analogous to the main text, we learn that the R\'enyi-$n$ SWSSB transition is described by a $d$-dimensional universality class for all $g > g_{c}$. Once more, the leading RG equations in the vicinity of the critical point are independent of the replica number $n$, and so we can analytically continue them to the replica limit $n \to 1$. We learn that a finite strength of decoherence is necessary to induce (R\'enyi-1) SWSSB\footnote{Notably, the behavior of R\'enyi-1 and R\'enyi-$n$ correlators (for example, $n = 2$) may be very qualitatively different close to the critical point. In particular, a ``replica ferromagnet'' phase is likely to arise for $n \geq 2$ in the vicinity of the critical point, leading to long-range order in $R_n(x,y)$ for $n \geq 2$ and short-range order for $R_1(x,y)$; see footnotes \cite{fn5} and \cite{fn6}.} even at the critical point $g = g_{c}$.

In the special case $d = 1$, the critical point of the TFIM can be described in terms of a free Majorana fermion CFT, and it can be shown that the inter-replica defect $\delta \mathcal{S}$ analogous to Eq.~\eqref{eq:S_defect} is \emph{exactly} irrelevant to all orders, so that no transition can occur until $\gamma = \infty$; see Ref.~\cite{weinsteinNonlocalityEntanglementMeasured2023a} for a very similar analysis. We therefore predict that the critical (1+1)$d$ TFIM under the decoherence channels \eqref{eq:swssb_decoherence} does not exhibit SWSSB until the maximal decoherence strength $p = \frac{1}{2}$, similar to the paramagnetic phase $g > g_{c}$.

\end{document}